# An advanced Study on Cryptography Mechanisms for Wireless Sensor Networks

Yassine MALEH
LAVETE Laboratory, Department of Computer Sciences
Faculty of Sciences and Technology, Morocco
y.maleh@uhp.ac.ma

Abdellah Ezzati
LAVETE Laboratory, Department of Computer Sciences
Hassan 1er University, Morocco
abdezzati@gmail.com

**Abstract**—Wireless Sensor Network (WSN) is consisting of independent and distributed sensors to monitor physical or environmental conditions, such as temperature, sound, pressure, etc. The most crucial and fundamental challenge facing WSN is security. Due to minimum capacity in-term of memory cost, processing and physical accessibility to sensors devices the security attacks are problematic. They are mostly deployed in open area, which expose them to different kinds of attacks. In this paper, we present an illustration of different attacks and vulnerabilities in WSN. Then we describe and analyze security requirement, countermeasures based on cryptography mechanisms in literature. Finally, we present possible directions in future research.

## I. Introduction

Inexpensiveness, energy efficiency, and consistency in performance is the need of the day for electronic communication. This has led to the advancement of wireless technologies and micro-electro-mechanical systems (MEMS). Which in turn made it possible to make low power tiny devices that runs autonomously. Which evolve a new class of distributed networking named wireless sensor networks (WSNs).

Today we find this kind of network in a wide range of potential applications, including security and surveillance, control, actuation and maintenance of complex systems and fine-grain monitoring of indoor and outdoor environments. The majority of these applications are deployed to monitor an area and to have a reaction when they register a critical event. The data does not need to be confidential in areas such as capturing indoor and outdoor environmental events. However, the confidentiality of data can be essential in other applications, such as for the security of a territory in military [1] [2].

Security is one of the most crucial challenge that WSN is facing. Sensor networks are highly vulnerable against attacks, it is very important to have certain mechanisms that can protect the network from all kinds of attack. It must be ensure that the system is protected from any kind of attacks. The most important tools that ensure security and its services are the cryptography mechanisms. In this work, we focus on study and analysis cryptography algorithms used in WSN.

This paper is organized as follows. In summary, the chapter makes the following contributions:
- ➢ It proposes threat models and security goals for WSNs.
- ➢ It identifies various possible attacks and vulnerabilities in WSN.
- ➢ It presents a detailed security analysis and comparison of all the major cryptography algorithms for WSN.
- ➢ It ends with a conclusion and future works.

## II. Security Goals for WSN

A sensor network is a special type of network. It shares some commonalities with a typical computer network, but also poses unique requirements. Therefore, we can think of the requirements of a wireless sensor network as encompassing both the typical network requirements and the unique requirements suited solely to wireless sensor networks. We can classify the security goals into two goals: main and secondary. The main goals include security objectives that should be available in any system (confidentiality, availability, integrity and authentication). The other category includes secondary goals (self-organization, secure localization, Time synchronization and Resilience to attacks) [2] [3].

### A. Data confidentiality

To ensure that the content of the message should not be revealed to the unauthorized receiver. Some secure data aggregation schemes provide this property in hop-by-hop, basis in which any aggregator node needs to decrypt the received encrypted data before applying the aggregate function on it, then encrypt the aggregated data before transmitting it to the higher-level aggregator or directly to the base station. While the other schemes provide end-to-end data confidentiality in which any aggregator node directly apply the aggregation function to the received encrypted data.

### B. Data integrity

Data integrity guarantees that the message has not been altered during the propagation. But if data aggregation is employed then it is not possible to have end-to-end data integrity since data aggregation yields in alteration. The data should be accessible only to authorized users.

*C. Source authentication*

Enables sensor node to ensure the identity of the peer node that it is communicating with. A compromised node can launch Sybil attack in which it may send data under several fake identities in order to corrupt the aggregated data.

*D. Availability*

To guarantee the survivability of network services against Denial-of-Service attacks. The attack aiming at an aggregator can make some part of the network losses its availability because the aggregator is responsible to provide the measurement of that network part.

*E. Data freshness*

It is also needed that the data is recent and no old message have been replayed. To achieve this goal a nonce, or another time related counter is added to the packet.

*F. Self Organization*

According to different situations a wireless sensor networks requires every sensor node to be independent and flexible enough to be self-organizing and self-healing. As it is a typically an ad hoc network. In sensor network there is no fixed infrastructure available for the purpose of network management, which make wireless sensor network security more challenging.

*G. Time synchronization*

Most of the sensor network applications depends on some form of time synchronization. Sensor may also require to computer end-to-end delay of a packet as it travels between pairwise sensors.

*H. Secure Localization*

The utility of a sensor network mostly rely on its ability that it automatically and accurately locate each sensor in the network. In order to point out the accurate fault a sensor network designed will need accurate location information. An attacker can easily exploit this situation and can easily manipulate non-secured location information by either reporting false signal strengths or replaying signals.

### III. DIFFERENT KIND OF ATTACKS AND VULNERABILITIES IN WIRELESS SENSOR NETWORKS

The different characteristics of wireless sensor networks (energy limited, low-power computing, use of radio waves, etc...) expose them to many security threats. The two general categories of attacks that are possible on a wireless network are active and passive attacks listening and monitoring the communication channel by unauthorized attackers is considered as passive attack. These attacks make it possible to retrieve data from the network but do not influence over its behavior. In active kind of attack the unauthorized attacker, neither only listen and monitor but also modifies the data in the communication channels. Unlike passive attacks, they directly hinder the provisioning of services. Active attacks can be classified into the following kinds. [4] [5] [6].

*A. Routing Attacks in sensor networks*

1. Spoofed, Altered, Or Replayed Routing Information

The most direct attack against a routing protocol is to target the routing information exchanged between nodes. By spoofing, altering, or replaying routing information, adversaries may be able to create routing loops, attract or repel network traffic, extend or shorten source routes, generate false error messages, and partition the network, increase end-to-end latency and so on.

2. Selective Forwarding

Only certain packets can be drop selectively by a malicious node. This is effective specially if is combined with an attack which gather much traffic through the node. It is assumed in sensor networks that nodes sincerely forward and receive messages. In situation where some of the compromised node refused to forward packet the neighbors may start using another route. [7] [8]

3. Spoofed, replayed and altered routing information

Ad hoc routing that is unprotected is vulnerable to these kind of attacks because every node act as a router and hence can directly affect routing information. Like:

- Extend or shorten service routes
- Generate false error messages
- Create routing loops
- Increase end-to-end latency

4. Sybil Attacks

This kind of attack targets fault tolerant schemes like multi path routing, distributed storage and topology maintenance. A node duplicate itself and present in the multiple locations. In the network a single node, present multiple identities in a Sybil attack as shown in figure 1. To prevent Sybil attack authentication and encryption techniques can be used. [7]

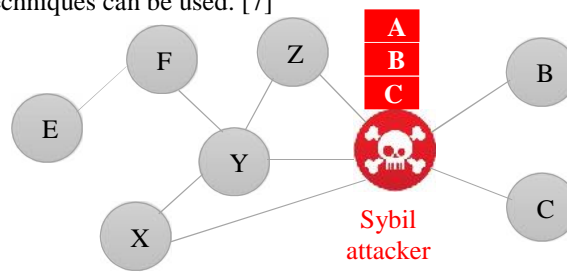

Fig. 1. Sybil attack

5. Sinkhole Attack

Sinkhole attack is to attract traffic to a specific node. The main goal of the attacker here is to attract nearly all the traffic from a particular area through a compromised node. This can be done by making the compromised node the most attractive to the neighbor nodes.

6. Wormhole Attack

In this kind of attack an attacker records packet at one location, tunnel them to another location in the network and then retransmit them into the network as shown in figure 2.

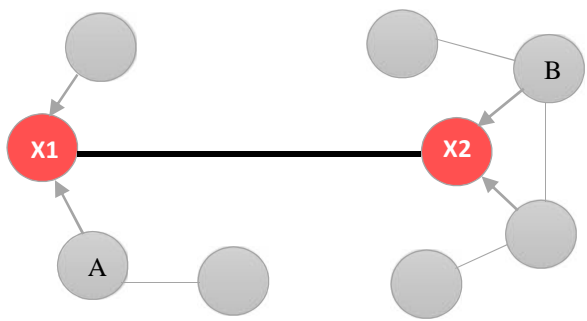

Fig. 2. Wormhole attack

7. Hello flood attack

In hello flood attack, the attacker uses hello packets as a weapon to convince the sensors in the network. Attacker sends a routing protocol hello packet from one node to another with more energy. An attacker with a high processing power and transmission range sends Hello packets to a number of sensor node that are isolated in a large area within the network. The sensors thus get convinced that the adversary is their neighbor. As a result, the victim nodes try to go through the attacker, as they know that it is their neighbor while sending the information to the base station (BS) as shown in figure 3.

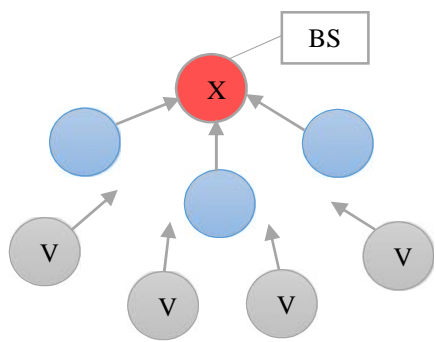

Fig. 3. Hello flood attack

*A. Denial of service Attack*

In this kind of attack, the objective of the malicious adversary is to avoid the provisioning of service. The most common and basic dos attacks can be:

1. Power exhaustion attack

An attacker impose a complex task to a sensor node in order to use its battery life. As we are aware of the fact sensor are low power devices with limited supply of energy. In addition, sensor nodes are limited with computational capabilities, so this attack can slow down the reaction time.

2. Jamming Attack

This is the primary physical layer DoS attack against WSN. In this kind of attack, the attacker emits radio frequency signals that do not follow an underlying MAC protocol, which lead to a problem that none of the member of the network in the affected area will be able to send or receive any packet. In other words, the adversary tires to transmit signals to the receiving antenna at the same frequency band or sub band as the transmitter, which causes radio interference. The attacker needs high energy to disrupt continuously the network. [9]

*B. Node subversion*

In this kind of attack, the whole sensor network may be compromised because the capture node may reveal cryptographic keys.

*C. Node Malfunction*

In this kind of attack a malfunction node, generate inaccurate data, which can risk the integrity of sensor network. This is more significant if the node is data aggregating node like cluster leader.

*D. Node Outage*

In this kind of attack a situation, occur where the node stops its function. Hence, the sensor network protocol should be robust enough to mitigate the effects of node outages by providing an alternate route.

*E. Physical Attacks*

As we have mentioned repeatedly that sensor network operate in outdoor environment mostly which make highly susceptible to physical attack. This code be physical node destruction. This attack is different from others in the sens that here a node can be permanently damaged or destroyed.

*F. False Node*

In this kind of attack a node is added by an adversary, which causes injection of false data. A node can be added to the system, which feeds false data or prevents the real data. This is the most dangerous attack cause a malicious code injected in the network could spread to all nodes which could potentially destroy the whole network. A more worse case could be talking over the network on behalf of an adversary.

*G. Node Replication Attacks*

In this kind of Attack a node is added to the existing sensor network by copying the node ID of an existing sensor node. This can severely disrupt the sensor network performance. The packet can be misrouted or even corrupted, which can lead to disconnected network or false sensor reading etc. Once the attacker get physical access to the entire network, it can copy the keys to the replicated sensor nodes.

*H. Passive information Gathering*

If the sensor network is not encrypted, an adversary having powerful resources can collect information from it. If the intruder is:

- Appropriate power receiver
- Well designed antenna

Then it can easily pick off the data stream. If the message interception contains the physical location of sensor nodes then it allows attacker to locate the node and destroy them. The adversary can observe the specific content of a message that include timestamps, message IDs and other fields. Strong encryption techniques needs to be used in order to prevent passive information gathering.

## IV. COUNTERMEASURES

Wireless sensor networks are highly vulnerable against attacks, it is very important to adopt some mechanisms that can protect the network from all kinds of attacks. It must be ensure that the system is protected before, during and after any kind of attack. The most important tools that ensure security and its services are the security primitives [10]. Those primitives are symmetric key encryption (SKE), public key cryptography (PKC) and hash functions [11]. SKE and Hash functions are the primitives that can be called the building blocks which offer a basic protection of the information flow, because these assure the confidentiality and integrity of the channel. Public key cryptography assure protection from the participation of external entities and also eliminate the problem of a malicious insider which try to use more than one identity. PKC assures this by allowing authentication of the peers involved in the information exchange. Based on the primitives it is possible to create a better network services. It is also equally important to have a key management system for constructing a secure key infrastructure.

### A. Methods and protocols classification

Most methods based on symmetric, asymmetric or hybrid systems solve the problem of key establishment through a predistribution phase. The predistribution of encryption keys in a WSN is the fact of storing these keys in the memory nodes before deployment. In literature, we find several classifications of cryptographic key management systems, such as papers in [12] - [14]. Some classifications methods are based on key sharing between two nodes (Pair-wise) or more nodes (Group-wise), and others rely on exploiting the probabilities, combinatory analysis, etc.

We chose to make a classification, which includes all key management and distribution models into two large families. The first family contains the asymmetrical schemes and the second includes the symmetrical schemes. Figure 4 illustrates this classification. In the following, we will detail the main models in literature.

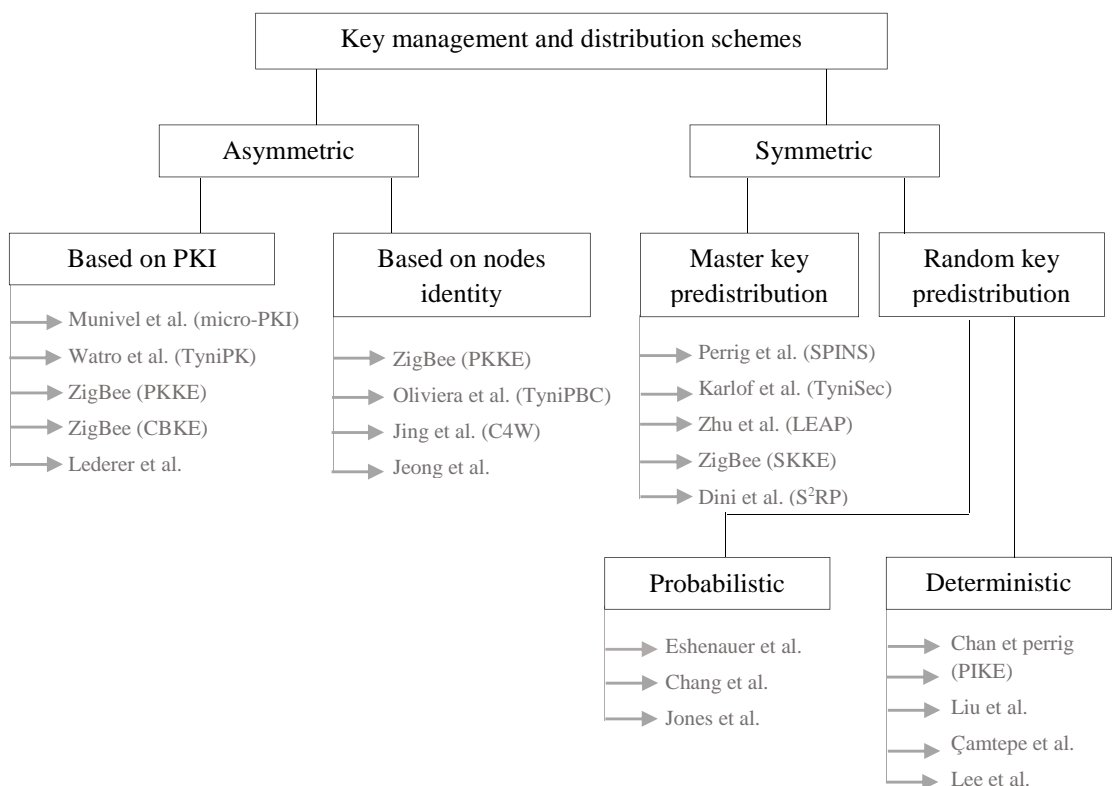

Fig. 4. Key management models

### B. Symmetric Schemes

The schemes in this category use symmetric mechanisms in order to establish a common key between two nodes in a WSN. This is accomplished in three steps:

- Key predistribution: keys stored in memory before deployment constitute the key ring of node. If there is a common key between two nodes, they can create a secure connection between them.
- Shared-key discovery: After deploying the communication protocol is responsible for discovering the common key between two neighboring nodes.
- Path-key establishment: if there is no common key between two nodes wishing to communicate, there must then find a secure path between them. This path goes through a set of nodes that already contains secure links. Once the path established, the two nodes can use it to secure communication.

We present in the following symmetrical schemes according to the decomposition of figure 4.

1. SPINS

SPINS is a suite of security building blocks proposed by Perig and several other authors in [15]. It is optimized for resource constrained environments and wireless communication. SPINS has two secure building blocks: SNEP and µTESLA. SNEP uses a shared counter between the two communicating parties and applies the counter in calculating encryption and a message authentication code (MAC) to provides data confidentiality, semantic security, data integrity, two-party data authentication, replay protection, and weak message freshness. What's more, the protocol also has low communication overhead, for the counter state is kept at each end point and the protocol only adds 8 bytes per message. For applications requiring strong freshness, the sender creates a random nonce (an unpredictable 64-bit value) and includes it in the request message to the receiver. The receiver generates the response message and includes the nonce in the MAC computation. µTESLA provides authenticated broadcast for severely resource-constrained environments. µTESLA constructs authenticated broadcast from symmetric primitives, but introduces asymmetry with delayed key disclosure and one-way function key chains. SPINS realizes an authenticated routing application and a security two-party key agreement with SNEP and µTESLA separately with low storage, calculation and communication consumption. However, SPINS still have some underlying problems as follows:

- It doesn't consider the possibility of DOS attack;
- Due to use the pairwise key pre-distribution scheme I the security routing protocol, SPINS rely on the base station excessively;
- SPINS doesn't consider the update of communication key. There must be practical key update mechanism to realize forward security;
- SPINS can't solve the problem of hidden channel leak and compromise node.

2. LEAP

LEAP (Localized Encryption and Authentication Protocol) is a key management protocol for sensor networks that is designed to support in-network processing with the prime goal at the same time to restrict the security impact of a node which is compromised to the immediate network neighborhood. The idea of leap+ was motivated after having this interesting observation that different types of messages that are exchanged between sensor nodes have different requirements of security. This observation gives the conclusion that a single keying mechanism is not suitable for meeting these different security requirements [16] [17] [18]. For each node leap+ support the establishment of four types of keys:

- Individual key: Shared with the base station;
- Pairwise key: Shared with another sensor node;
- Cluster Key: Shared with multiple neighboring nodes;
- Global key: Shared by all nodes in the network.

The packets that each node exchanged in a sensor network can be classified into several categories, which is based on different criteria for example:
- Control packets vs Data packets
- Broadcast Packets Vs Unicast Packets
- Queries or commands Vs Sensor readings and so on.

The security requirement for each packet is different it depends on the category it falls in. Almost all type of packets requires authentication while confidentiality is only for some types of packets. Here it is mentioned that single keying mechanism is not appropriate for all the secure communication that are needed in sensor networks.

3. Tinysec

Karlof et al. [19] propose the TinySec Protocol, the first full implementation of a secure architecture at the data link layer for WSN. This implementation supports two security options: a message authentication with data encryption (TinySec-EA) and authentication of messages without data encryption (TinySec-Auth). As SPINS, TinySec uses standard cryptographic algorithms to ensure privacy and message integrity check. The authors of Tinysec find that Skipjack algorithm [20] is more suitable for WSN than RC5 (algorithm used by SPINS). Indeed, evaluations of TinySec have shown that RC5 needs a pre-key calculation using 104 bytes of RAM. TinySec uses the CBC encryption mode (Cipher Block Chaining) instead of the CTR (used by SPINS). Indeed, the CTR will provide for more packet encryption the same random numbers. These numbers are used primarily in the production of the encryption keys sequences, their repetition can weaken the security level of this solution and subsequently allowing adversaries to discover the content of messages. TinySec is an implementation rather than a key distribution proposal, he comes to complete a key distribution method suited to the expanded network. Two nodes need two shared symmetric key to communicate. The first is used to encrypt messages and the second for calculating MAC (code) messages.

*C. Asymmetric schemes*

The schemes in this category use the mechanisms of asymmetric systems in order to establish a common key between two nodes or a group of nodes of a WSN.

1. micro-PKI

Munivel et al. [21] propose a method for WSN called micro-PKI (Public Key Infrastructure Micro), a simplified version of conventional PKI. The base station has a public key and another private. The public key is used by the network nodes to authenticate the base station, and the private key is used by the base station to decrypt data sent from the nodes. Before deployment, the public key of the base station is stored in all nodes. The authors include in their scheme two types of authentication (HandShake). The first type of authentication occurs between a network node and the base station. The node generates a symmetric session key and encrypts it with the public key of the base station. To ensure the integrity of messages exchanged, the authors propose to integrate with each message a MAC (code) using the same encryption key of the message. For new nodes who wish to join the network, they simply store in these nodes, the public key of the base station before deployment.

2. TinyPK

Watro et al. [22] proposed a method called TinyPK based on the use of public keys and the principle of Diffie-Hellman to establish a secret key between two nodes in a WSN. TinyPK uses a trusted authority to sign the public keys of nodes. The CA key is predistributed to all nodes before deployment so they can check key neighbors after deployment. The choice of the RSA algorithm for encryption involves a great consumption of time and energy of the nodes. Thus, the basic operations can take dozen seconds, which will reduce the network lifetime as well as impact on reactivity.

3. PKKE & CBKE

The PKKE and CBKE protocols proposed by Zigbee using the identity of nodes in their method of key establishment. The goal is to use these identities to create a single shared key between each pair of nodes in a network. However, the creation of the shared key is performed with interactions between the two nodes. It means, methods require sending and receiving multiple messages on both sides before the creation of the key. To save power nodes that want to share a secret and those intermediate nodes, several methods have been proposed to remove these interactions. These methods are known in the field of cryptography as the ID-NIKDS [23] (Identity-Based Non-Interactive Key Distribution Scheme).

4. C4W

Jing et al. [24] proposed a method called C4W based on the use of the identity of nodes to calculate public keys. The nodes themselves are able to calculate the public keys of other nodes using their identities. What could replace the role of a certificate. Before deployment, the nodes and the base station are loaded with their own keys (private / public key ECC) and public information on the network nodes. The C4W method uses the principle of Diffie-Hellman key exchange to create a single shared key between two nodes without using certificates

IV. Analysis and discussion

A. Analyses method

Several criteria are taken into account in order to compare the different methods for key management. We present in Figure 5 the most important criteria. We begin by the limitation of the resources of nodes. The proposed key management method must consider the fact that the nodes have been deployed to collect the information. They need their memory space to store their data and their embedded energy to ensure their application role. The solution must also be flexible and dynamic, and able to go to the scale (scalability). Another criterion, which must be respected, is the resilience against attacks. When capturing nodes for example, the opponent can use the information stored to implement other attacks and manipulate the network. The key management method must be able to detect compromised nodes and authenticate network nodes before distributing the keys. The last criterion is the renewal and revocation of keys. We can put it at the same level of importance as key distribution. An expired key or discovered by an opponent must be revoked. The keys secure links also need to be renewed periodically. The connectivity of a network is the guarantee of its nodes to have more secure paths to send its data. The method of keys distribution must be capable of ensuring a good network connectivity. The case of a departure or a capture of a node may limit the connectivity of other nodes with the network. The distribution method must take into account this factor by proposing new secure paths.

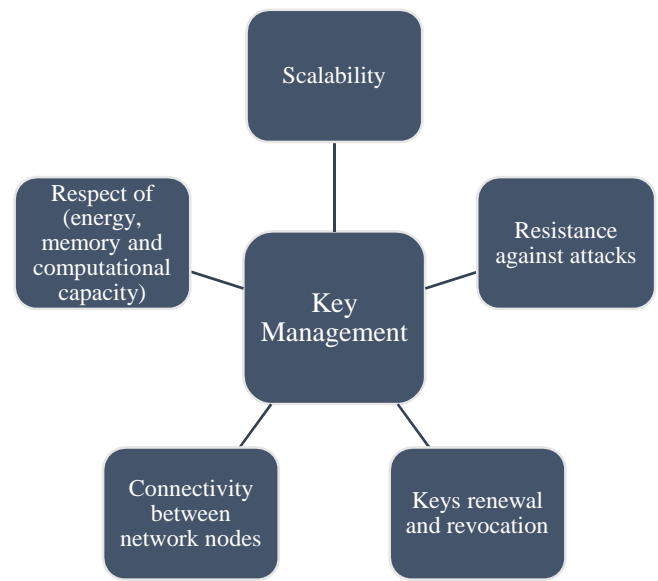

Fig. 5. The criteria for comparison of methods of key management of RCSF

B. Comparison and discussion

We have studied different types of distribution and key establishment in the WSN and we have class the diagrams of these different types in figure 4. In table I, we compare these diagrams based on the criteria of comparison in figure 5. Note that the memory storage evaluated in the table takes into account only the size of keys stored in the nodes and not the size of the code algorithms and cryptographic primitives. The stars in the table shall designate a quality. We have established three, two or one black star in the column "Connectivity" to note that the diagrams have respectively a high, medium or low connectivity. Then we have established three, two or one star in the column "Resistance against the attacks "to note that the diagrams have respectively a high, medium or low resistance against attacks. On the other hand, the circle in the table shall designate a default. We have three, two or one black circle in the column "Cost resources used" to note that the diagrams respectively cost a high, medium and low-resource utilization.

| Schemes | | | Criteria of comparison | | | | | | | | |
|---|---|---|---|---|---|---|---|---|---|---|---|
| Type | Authors | | Scalability | Connectivity | Resistance against attacks | | | | Resource cost used | | |
| | | | | | Information collection | Communication perturbation | Data aggregation and resource exhausted | Capture of physical nodes | Memory (key store) | Calculation and energy consumption | Renewal and revocation |
| Symmetric key schemes | *Chan et al.* [26] | Proba. | Limited | ★★☆ | | | | | ●●● | ●●○ | |
| | *Chan et Perrig (PIKE)* [27] | Deter. | No | ★★★ | | ★☆☆ | ★☆☆ | | | ●○○ | ●●● |
| | *Perrig et al.(SPINS)* [15] | MK + BS | Limited | ★★☆ | | | | | ●○○ | ●●○ | |
| | *Zhu et al.(LEAP)* [16] | MK | Good | ★★☆ | | | | | | | |
| Public key schemes | *ZigBee (PKKE)* [28] | ID | Good | ★☆☆ | ★★★ | | | | ●●● | ●●○ | |
| | *Munivel et al.(micro-PKI)* [21] | PKI | Limited | ★☆☆ | | ★★☆ | ★★☆ | | ●○○ | ●●● | ●○○ |
| | *Watro et al. (TinyPK)* [22] | PKI | Limited | ★☆☆ | | | | | | | ●●● |

Table 1. Comparison of the proposed key management schemes for WSN. In this table circle is used to indicate a fault, while the star designate a quality.

The diagrams SPINS and LEAP use master keys in the key establishment. This reduces the storage of keys in the memory of the nodes. However, the resistance to attacks is low. Given that the master key can be compromised at any time, the keys established after the deployment by using this key can be compromised also. By adopting a symmetric system, they are the most suitable and among the most rapid in terms of calculation. Note that the symmetric diagrams are costly in operations (if they exist) of renewal and revocation of keys since they use secret keys in order to exchange other secret keys. The problem is simpler in the asymmetric diagrams since the public keys do not need to be secret.

The scheme Chan et al., Representing probabilistic diagrams shows that consumes low power and do not require much computing capacity. However, large sizes key rings stored in the memory nodes before deployment makes this scheme one of the most expensive symmetrical schemes in terms of memory occupation. It cannot resist to attacks of type physical nodes captures. While PIKE scheme provides better connectivity between network nodes, but it shows low performance in term of scalability.

We can see that the schema TinyPBC is the most suitable of asymmetric patterns. It is resistant to most known attacks in the RCSF. The fact of using the coupling in order to establish a unique key shared between two nodes has helped reduce the need for large storage capacity in memory. In addition, the creation of this key is performed without interaction between the nodes which saves the time of calculation and the energy consumed due to these interactions. The diagrams using the principle of certificates and PKI remain the most expensive in calculation and in energy consumption.

The comparison between the diagrams symmetrical and asymmetrical may differ depending on the desired level of security in the network. We note in the table of comparison that the symmetric diagrams can be chosen for their timeliness and the asymmetric diagrams for their resistance against attacks.

## V. Conclusion

We have presented in this paper, a state of the art, which detailed the attacks threatening the WSN. To face those attacks, we have presented a synthesis of cryptography systems and mechanisms that can secure the WSN. The lack of infrastructure such as PKI in the WSN has compelled the nodes to not have confidence in network and to create secure paths from the source of the data to the base station. Works like [29] have used the identities of the nodes and the principle of coupling in order to reduce, or even eliminate, the interactions between nodes to counter a maximum of attacks. However, to date there is no complete and dynamic solutions easily adaptable to WSN. All the methods we have discussed in this paper have great advantages when dedicated to specific applications and network topologies. However, it is difficult if not impossible to adapt them to other kinds of applications or make them polyvalent and universal.

The current secure routing protocols such as LEAP and SPINS are mature in the routing performance. In the design of secure routing protocol, if through a small amount of change we can achieve security purposes, the expansion of the original protocol can also offers a good research method. Such as SRD, INTRSN, etc. Furthermore, this direction is our research direction in the future.